\documentstyle[12pt]{article}  
\title{Electromagnetic properties of\\ 
 neutral and charged spin 1 particles}

\author{Jos\'e F. Nieves\\ 
Laboratory of Theoretical Phsyics\\
Department of Physics, University of Puerto Rico\\
P.O. Box 23343, 
R\'{\i}o Piedras, Puerto Rico 00931-3343\\
\and
Palash B. Pal\\ 
Indian Institute of Astrophysics, Bangalore 560034, India}

\date{hep-ph/9611431}

\begin{document}

\maketitle 

\begin{abstract}                               
The structure of the electromagnetic vertex of spin-1 particles are
studied in a general way, for the diagonal as well
as the off-diagonal couplings.
In each case, we consider in detail the
consequences of gauge invariance and the space-time
discrete symmetries, paying particular attention
to the implications for
the couplings of the photon to neutral gauge bosons, which arise in
higher orders.  Several general results concerning
the static electromagnetic properties of neutral bosons
are derived, analogous to the results that are well known
in the case of neutral spin-1/2 particles. 

\end{abstract}

%
%
\section{Introduction and Notation}\setcounter{equation}{0}\label{sec:introduction}
The properties of vector bosons has been the subject 
of intense studies recently\cite{reviews}. This is 
motivated mainly by the fact
that direct measurement of the trilinear vector boson
couplings may be possible in the pair production
of electroweak bosons at $e^+e^-$ and hadron
colliders. A possible way to test whether the predictions of the
Standard Model are sustained is to parametrize
the trilinear vector boson couplings in some
generalized form, and then try to constrain
the generalized couplings from the experimental results. 
A consistent implementation of 
such parametrization must satisfy certain requirements
that follow from fundamental principles such
as Lorentz and gauge invariance, the discrete
space-time symmetries C, P, T, and their combinations, or any other
symmetry that may be applicable.
This has been done in great detail for the 
$W^+W^-\gamma$ vertex \cite{HPHZ87,WWgamma}, 
which awaits experimental tests in the near future.

Although neutral gauge bosons cannot have a coupling with the photon at the 
tree level, they do develop such couplings in higher orders
of perturbation theory. 
Thus, for example, the standard 
model admits a $ZZ\gamma$ coupling at loop levels. 
It is conceivable that vector bosons in addition
to the standard $W$ and $Z$
exist in Nature.  In fact, it is not uncommon to
find in the literature 
theories beyond the standard model that predict
the existence of such additional bosons.  In those theories,
the trilinear couplings involving the photon
can be more complicated than those of the Standard Model.
In particular, for example,
off-diagonal couplings $VV^\prime\gamma$ involving different 
vector bosons can in principle exist.  
While such couplings arise only at loop levels, they may be important for
discovering physics beyond the standard model. For example, a heavy
gauge boson may be detected by its radiative decay into a
lighter one.  In order to prepare ourselves for
these possibilities, 
it is useful to look in a
more general way at the electromagnetic vertex of neutral 
spin-1 particles. This is our purpose in this paper.
We study the structure of 
the electromagnetic vertex of charged as well as neutral
spin-1 particles, for the diagonal and
for the off-diagonal couplings. We consider in detail the
consequences of gauge invariance and the space-time
discrete symmetries, paying particular attention
to the implications for
the electromagnetic couplings of neutral gauge bosons.
This kind of study can serve as a guide
in the quest to find the physics that may lie beyond
the standard model in a way that is
general and model-independent.  Moreover,
it may allow us to recognize possible
deviations of the new physics from
basic physical principles such as
Lorentz and gauge invariance, and crossing symmetry.

We begin by introducing the notation for the
electromagnetic vertex of spin-1 particles.
The off-shell vertex function $\Gamma_{\alpha\alpha'\mu}(k,k')$
is defined such
that the matrix element of the electromagnetic current
is
\begin{eqnarray}\label{defGamma}
\langle V'(k')|j_\mu^{(EM)}(0)|V(k)\rangle
& = & \epsilon^{\prime\ast\alpha'}(k')\epsilon^\alpha(k)
\Gamma_{\alpha\alpha'\mu}(k,k')\nonumber\\*
& \equiv & j_\mu(Q,q)  \,,
\end{eqnarray}
corresponding to the vertex
\begin{equation}
\label{V->V'A}
V(k) \rightarrow V'(k') + \gamma(q)\,.
\end{equation}
For the sake of convenience, we represent the function 
$j_\mu$ in terms of the two independent momenta
\begin{eqnarray}
Q \equiv k+k', \qquad q \equiv k-k' \,,
\label{defQq}
\end{eqnarray}
where $q$ is the photon momentum as depicted in Eq.\ (\ref{V->V'A}).

Before we write the
most general form of the vertex consistent with Lorentz
invariance, it is useful to make the following observations:
(i) For $V$ and $V'$ are neutral, electromagnetic gauge 
invariance implies
\begin{equation}\label{gaugeinvstrong}
q^\mu\Gamma_{\alpha\alpha'\mu} = 0\,,
\end{equation}
for arbitrary values of $k$ and $k'$, and hence of $q$.
On the other hand, if $V$ and $V'$ are charged,
the relation analogous to Eq.~(\ref{gaugeinvstrong}) contains
some terms in the right-hand side involving the
inverse propagators of $V$ and $V'$.
Therefore, what we get is the weaker condition
\begin{equation}\label{gaugeinvweak}
q^\mu j_\mu(Q,q) = 0\,,
\end{equation}
where $j_\mu(Q,q)$ is the on-shell vertex function as defined
in Eq.~(\ref{defGamma}),  which
follows simply by using the current conservation
condition 
\begin{equation}\label{currcons}
q^\mu \left< V' (k') \left| j_\mu^{(EM)}(0) \right| V(k) \right> = 0
\end{equation}
in Eq.~(\ref{defGamma}).  
(ii) For neutral $V$ and $V'$, $\Gamma_{\alpha\alpha'\mu}$ 
is well defined in the limit $q\rightarrow 0$. 
This property does not hold for $V$ and $V'$ charged.
(iii) Ultimately, for the calculation
of an amplitude for a physical process, the indices
$\alpha',\alpha$ of $\Gamma$ will be
contracted with the polarization vectors 
$\epsilon^{\prime\ast\alpha'}\epsilon^\alpha$ of $V'$ and $V$,
or with fermion currents, and similarly for the photon index $\mu$.  
%
%
%
%
\section{Parametrization of the vertex and its physical interpretation}\setcounter{equation}{0}
\label{sec:parametrization}
Let us consider for the moment the amplitude for processes in which the vector
particles are on-shell and/or the 
fermion currents to which they couple are conserved. From the 
transversality conditions
\begin{equation}\label{transversality}
k'\cdot\epsilon' = k\cdot\epsilon = 0 \,,
\end{equation}
we obtain
\begin{eqnarray}\label{polvecrelations}
Q\cdot\epsilon  =  - q\cdot\epsilon\,, \qquad 
Q\cdot\epsilon'  =  \phantom{-}  q\cdot\epsilon'\,.
\end{eqnarray}
Similar relations
to Eq.\ (\ref{polvecrelations}) hold for the fermion
currents if they are conserved, or in the limit in which
the fermion masses can be neglected.
Therefore the terms in $\Gamma_{\alpha\alpha'\mu}$
proportional to $Q_\alpha$
or $Q_{\alpha'}$ are not independent of similar terms
with $Q_\alpha$ replaced by $q_\alpha$ and 
$Q_{\alpha'}$ replaced by $q_{\alpha'}$.
In this case
the most general form of $\Gamma_{\alpha\alpha'\mu}$
that is consistent with Lorentz invariance can be written 
in terms of 10 form factors as follows:
\begin{eqnarray}\label{Gammagen}
\Gamma^{(T)}_{\alpha\alpha'\mu} & = & 
(a_1 q_\mu + a'_1 Q_\mu) g_{\alpha\alpha'} +
(a_2 q_\mu + a'_2 Q_\mu) q_\alpha q_{\alpha'} +
a_3 (g_{\mu\alpha'}q_\alpha - g_{\mu\alpha}q_{\alpha'}) +
a_4 (g_{\mu\alpha'}q_\alpha + g_{\mu\alpha}q_{\alpha'})\nonumber\\* 
& & \mbox{} +
b_1\varepsilon_{\alpha\alpha'\mu\nu}q^\nu + 
b_1'\varepsilon_{\alpha\alpha'\mu\nu}Q^\nu +
b_2 q_\alpha [Qq]_{\mu\alpha'} + 
b_3 q_{\alpha'} [Qq]_{\mu\alpha} 
\end{eqnarray}
where we have defined
\begin{equation}\label{defbracket}
[Qq]_{\alpha\alpha'} \equiv 
\varepsilon_{\alpha\alpha'\beta\gamma}Q^\beta q^\gamma
\end{equation}
for the sake of brevity. 
For future reference, we note here that our convention
for $\varepsilon_{\alpha\alpha'\beta\gamma}$ is such that
\begin{equation}
\varepsilon^{0123} = +1 \,.
\end{equation}
In principle Eq.~(\ref{Gammagen}) can contain the terms
\begin{equation}\label{hterms} 
(h_1 q_\mu + h_2 Q_\mu) [Qq]_{\alpha\alpha'}\,,
\end{equation}
but in fact these are redundant.  
To see this, recall the identity
\begin{equation}\label{epsilonident}
g_{\lambda\mu}\varepsilon_{\alpha\alpha'\beta\gamma} -
g_{\lambda\alpha}\varepsilon_{\mu\alpha^\prime\beta\gamma} -
g_{\lambda\alpha'}\varepsilon_{\alpha\mu\beta\gamma} -
g_{\lambda\beta}\varepsilon_{\alpha\alpha'\mu\gamma} -
g_{\lambda\gamma}\varepsilon_{\alpha\alpha'\beta\mu} = 0\,,
\end{equation}
which follows from the fact that the left-hand side is a
completely antisymmetric four-dimensional
tensor in the indices $\mu\alpha\alpha'\beta\gamma$.
Contracting Eq.~(\ref{epsilonident}) 
with $q^\lambda Q^\beta q^\gamma$, we obtain
\begin{eqnarray}\label{epsq}
q_\mu [Qq]_{\alpha\alpha'} = 
-q^2\varepsilon_{\alpha\alpha'\mu\nu}Q^\nu + 
Q \cdot q \, \varepsilon_{\alpha\alpha'\mu\nu} q^\nu -
q_{\alpha'} [Qq]_{\mu\alpha} +
q_\alpha [Qq]_{\mu\alpha'} \,. 
\end{eqnarray}
Similarly, by contracting  Eq.~(\ref{epsilonident}) 
with $Q^\lambda Q^\beta q^\gamma$, we obtain
\begin{eqnarray}\label{epsQ}
Q_\mu [Qq]_{\alpha\alpha'} =
Q^2\varepsilon_{\alpha\alpha'\mu\nu}q^\nu -
Q \cdot q \, \varepsilon_{\alpha\alpha'\mu\nu} Q^\nu -
Q_{\alpha'} [Qq]_{\mu\alpha} +
Q_\alpha [Qq]_{\mu\alpha'} \,. 
\end{eqnarray}
In the last two terms in this equation, 
$Q_{\alpha'}$ and $Q_\alpha$ can be replaced by 
$q_{\alpha'}$ and $q_\alpha$ 
due to Eq.~(\ref{polvecrelations}),
which finally justifies the
omission of $h_{1,2}$ in Eq.~(\ref{Gammagen}).

On the other hand, for amplitudes
in which the vector particles are not on-shell
and the fermion currents to which they couple
are not conserved, the simplification made
by using Eq.~(\ref{polvecrelations}) is not valid.
In these cases, the general form of the vertex function 
contains in addition to the terms given in Eq.~(\ref{Gammagen})
the following:
\begin{eqnarray}\label{Gammagen2}
\Gamma^{(L)}_{\alpha\alpha'\mu} & = &
(c_1 q_\mu + c'_1 Q_\mu) k_\alpha k'_{\alpha'} +
(c_2 q_\mu + c'_2 Q_\mu) k_\alpha k_{\alpha'} +
(c_3 q_\mu + c'_3 Q_\mu) k'_\alpha k'_{\alpha'} \nonumber\\*
& & 
+ c_4 g_{\mu\alpha'}k_\alpha + c_5 g_{\mu\alpha}k'_{\alpha'} +
d_1 k_\alpha [Qq]_{\mu\alpha'} + 
d_2 k'_{\alpha'} [Qq]_{\mu\alpha} \,.
\end{eqnarray}
In the most general case, the vertex function
is the sum
\begin{equation}\label{Gammatot}
\Gamma_{\alpha\alpha'\mu} =
\Gamma^{(T)}_{\alpha\alpha'\mu} + 
\Gamma^{(L)}_{\alpha\alpha'\mu}\,.
\end{equation}

For $V=V'$, 
the form factors in Eq.~(\ref{Gammagen}) have a 
definite physical interpretation in terms
of the static electromagnetic properties
of the particles.  To deduce this correspondence
we use the following procedure.
Let us consider as an example the electric dipole moment $d_E$.
The classical definition of the electric dipole
moment of a charge distribution is
\begin{equation}\label{DsubE}
\vec D_E = \int d^3\!x\,\vec x \rho^{(EM)}(\vec x)\,,
\end{equation}
where we have denoted by $\rho^{(EM)}(\vec x)$ the zeroth
component of the electromagnetic current density
$j^{(EM)}(\vec x)$.
In Quantum Mechanics $\vec D_E$ becomes an operator,
and the electric dipole moment of the particle
is given by its expectation value.  For normalizable
states, with $\langle ks^\prime|ks\rangle = N_k\delta_{s,s^\prime}$,
we can define the electric dipole moment matrix elements
as
\begin{equation}\label{dsubEdef1}
\xi^{\prime\dagger}\vec d_E \xi = 
(\vec d_E)_{ss^\prime} = \left. \frac{\langle ks^\prime|\vec D_E|ks\rangle}
{N_k}\right|_{\vec k\rightarrow 0}\,.
\end{equation}
The limit
$\vec k = 0$ reflects the fact
that the intrinsic static properties
of the particle are determined in the
limit of zero momentum or, equivalently, 
in the rest frame of the particle.
In Eq.~(\ref{dsubEdef1}) the symbols $\xi,\xi^\prime$ stand for
the space part of the polarization vectors
\begin{eqnarray}\label{spinwavefn}
\epsilon^\mu(k,s) & \equiv & (\epsilon^0,\vec\xi)\nonumber\\
\epsilon^{\prime \mu}(k^\prime,s^\prime) 
& \equiv & (\epsilon^{\prime 0},\vec\xi^{\, \prime})\,,
\end{eqnarray}
written as column matrices. In the limit $\vec k = 0$,
they are simply the spin wave functions of the particle
at rest.
However, since the plane wave states satisfy
\begin{equation}\label{planewavenorm}
\langle V(k^\prime s^\prime)|V(ks)\rangle = (2\pi)^3 2k^0 
\delta^{(3)}(\vec k - \vec k^\prime)\delta_{s,s^\prime}\,,
\end{equation}
the appropriate formula is
\begin{equation}\label{dsubEdef}
(\xi^{\prime\dagger}\vec d_E \xi)(2\pi)^3 2k^0\delta^{(3)}(\vec k - \vec k^\prime)
= \left. \langle V(k^\prime s^\prime)|\vec D_E|V(ks)\rangle
\right|_{|\vec k^{\prime}| = |\vec k|\rightarrow 0}\,.
\end{equation}
By translation invariance,
\begin{eqnarray}
\langle V(k^\prime s^\prime)|\vec D_E|V(ks)\rangle
& = & \int d^3\! x\, \vec x e^{i\vec x\cdot(\vec k - \vec k^\prime)}
\langle V(k^\prime s^\prime)|\rho^{(EM)}(\vec 0)|V(ks)\rangle\nonumber\\
& = & (2\pi)^3\delta^{(3)}(\vec k - \vec k^\prime)
\left[\frac{i\partial j^0(Q,q)}{\partial\vec q}\right]\,,
\end{eqnarray}
where, to arrive at the second equality, Eq.~(\ref{defGamma}) has been used.  
Then,
taking the limits indicated in Eq.~(\ref{dsubEdef}),
\begin{equation}\label{dsubEexp}
(\xi^{\prime\dagger}\vec d_E \xi) = \frac{1}{2m_V}
\left. \left[\frac{i\partial J^0(\vec q)}{\partial\vec q}\right]
\right|_{\vec q = 0}\,,
\end{equation}
where we have introduced
\begin{eqnarray}\label{bigJ}
J_\mu(\vec q) & = & \left. j_\mu(Q,q)\right|_{q^0 = 0,\vec Q = 0}\nonumber\\
& = &  \left.\left[\epsilon^{\prime\ast\alpha'}(k')\epsilon^\alpha(k)
\Gamma_{\alpha\alpha^\prime\mu}(k,k')\right]\right|_{q^0 = 0,\vec Q = 0}\,.
\end{eqnarray}
The expression for $\vec d_E$ can finally
be obtained by substituting Eq.~(\ref{Gammagen})
into Eq.~(\ref{dsubEexp}).
The terms contained in $\Gamma^{(L)}_{\alpha\alpha'\mu}$
do not contribute because
they vanish when the vertex function is multiplied
by the polarization vectors, and  it is important also
to keep in mind that, in the limit $q^0 = 0$ and $\vec Q = 0$,
\begin{eqnarray}\label{wavefnrelations}
\epsilon^0 & = & \frac{\vec q\cdot\vec\xi}{2m_V}\nonumber\\
\epsilon^{\prime 0} & = & -\frac{\vec q\cdot\vec\xi^\prime}{2m_V}\,,
\end{eqnarray}
which follow from Eq.~(\ref{transversality}). In this way we then
obtain
\begin{equation}\label{dsubEmatrix}
\vec d_E = d_E\vec S\,,
\end{equation}
where
\begin{equation}\label{dsubE}
d_E = 
\left(\frac{b_1(0)}{2m_V}\right)
\end{equation}
and we have introduced the spin-1 matrices $\vec S$ with elements
\begin{equation}\label{spinmatrices}
(S^i)_{jk} = -i\varepsilon^{ijk}\,.
\end{equation}
The notation $b_1(0)$ is meant to indicate that
the form factor is to be evaluated by 
first putting $k$ and $k^\prime$ on shell,
and then taking the limit $q^0 = 0$, $\vec Q = 0$ and
$\vec q = 0$.
Since, in vacuum, the on-shell form factors are functions only 
of $q^2$, in order to evaluate
$b_1(0)$ it is enough to put $q^2 = 0$.
However in more general situations, such as if the
particle is propagating in a matter background,
the form factors depend on the components of
$q^\mu$ and $Q^\mu$ separately,
and the static limit is meant to be
as we have indicated, namely, $(q^0 = 0,\vec q\rightarrow 0)$ and
$Q^\mu\rightarrow (2m_V,\vec 0)$.  In what follows, we use
the same notation for the other form factors.

We can proceed in similar fashion for the electric
charge and the quadrupole moment.  
The formulas analagous to Eq.~(\ref{dsubEdef}) are
\begin{eqnarray}\label{qsubEdef}
(\xi^{\prime\dagger} q_V \xi)(2\pi)^3 2k^0\delta^{(3)}(\vec k - \vec k^\prime)
& = & \langle V(k^\prime s^\prime)|q_E|V(ks)\rangle|_{|\vec k^{\prime}| = 
|\vec k|\rightarrow 0}\\
\nonumber\\
\label{XsubEdef}
(\xi^{\prime\dagger} x_E^{ij} \xi)(2\pi)^3 2k^0\delta^{(3)}(\vec k - \vec k^\prime)
& = & \langle V(k^\prime s^\prime)|X_E^{ij}|V(ks)\rangle|_{|\vec k^{\prime}| 
= |\vec k|\rightarrow 0}\,,
\end{eqnarray}
where $q_E$ is the charge operator
\begin{equation}\label{qopsubE}
q_E = \int d^3\!x\, \rho^{(EM)}(\vec x)\,.
\end{equation}
and 
\begin{equation}\label{XsubE}
X_E^{ij} = \int d^3\!x\, x^i x^j \rho^{(EM)}(\vec x)\,
\end{equation}
are the elements of the quadrupole moment tensor operator.
Following the steps leading to Eq.~(\ref{dsubEexp}), we obtain 
for $q_V$ and $x_E$
\begin{eqnarray}\label{qsubVexp}
(\xi^{\prime\dagger} q_V \xi) & = & \frac{1}{2m_V} J^0(\vec 0)\,,\\
\label{xsubEexp}
(\xi^{\prime\dagger} x_E^{ij} \xi) & = & \frac{1}{2m_V}
\left. \left[\frac{i\partial}{\partial q^i}
\frac{i\partial}{\partial q^j}J^0(\vec q)\right]
\right|_{\vec q = 0}\,,
\end{eqnarray}
from which we obtain
\begin{eqnarray}\label{Echarge}
q_V & = & -a_1^\prime(0)\,,\\
\label{EQM}
x_E^{ij} & = & \frac{1}{3}Q_E I^{ij}\,.
\end{eqnarray}
In Eq.~(\ref{EQM}) the $I^{ij}$ are a set of symmetric matrices
with elements
\begin{equation}\label{Isupij}
(I^{ij})_{kl} = \delta^i_k\delta^j_l + \delta^j_k\delta^i_l\,.
\end{equation}
and
\begin{equation}\label{QsubE}
Q_E = -3\left[a_2^\prime(0) - 
\frac{(a_1^\prime(0) + 2a_3(0))}{4m_V^2}\right]
\end{equation}
It is customary to define the quadrupole moment 
tensor as\cite{jackson}
\begin{eqnarray}
Q_E^{ij} & \equiv & 3x_E^{ij} - \delta^i_j\left(\sum_k x_E^{kk}\right)
\nonumber\\
& = & Q_E\left(I^{ij} - \frac{2}{3}\delta^i_j{\bf 1}\right)\,,
\end{eqnarray}
where {\bf 1} is stands for the $3\times 3$ unit identity matrix.

At this point it is useful to consider
the amplitude
for the scattering of a slowly moving $V$ particle in
an external electromagnetic field.  Introducing the potentials
\begin{eqnarray}\label{extpotentials}
\phi(\vec x) & = & \int \frac{d^3\vec q}{(2\pi)^3}e^{-i\vec q\cdot 
\vec x}\phi(\vec q)\nonumber\\
\vec A(\vec x) & = & \int \frac{d^3\vec q}{(2\pi)^3}e^{-i\vec q\cdot 
\vec x}\vec A(\vec q)\,,
\end{eqnarray}
The amplitude is
\begin{equation}\label{ampextfield}
iM = -i(\phi(\vec q)J^0(\vec q) - \vec A(\vec q)\cdot \vec J(\vec q))\,.
\end{equation}
If we expand $J^0(\vec q)$ in powers of $\vec q$,
the term in Eq.~(\ref{ampextfield})
that is independent of $\vec q$ is
then identified with the charge, those proportional
to $\vec q$ are identified with the dipole moments,
while those proportional to two powers of $\vec q$
correspond to the quadrupole moments. 
More precisely, from Eqs.\ (\ref{dsubEexp}), 
(\ref{qsubVexp}) and (\ref{xsubEexp})
we deduce that
\begin{equation}\label{J0A0}
\phi(\vec q)J^0(\vec q) = 
\xi^{\prime\dagger}(q_V\phi - \vec d_E\cdot\vec E
-\frac{1}{2}x_E^{ij}E^{ij})\xi\,,
\end{equation}
and, by analogy, we can determine the magnetic moments
by writing 
the amplitude in the form
\begin{equation}\label{ampextfieldform}
M  = -2m_V \xi^{\prime\dagger}(q_V\phi - \vec d_E\cdot\vec E
-\frac{1}{2}x_E^{ij}E^{ij}
- \vec d_M\cdot\vec B -\frac{1}{2}x_M^{ij}B^{ij})\xi\,.
\end{equation}
Here we have introduced
\begin{eqnarray}\label{extfields}
\vec E & = & i\vec q \phi\nonumber\\
\vec B & = & -i\vec q\times\vec A\,,
\end{eqnarray}
which are the Fourier transforms of the external electric
and magnetic field, and
\begin{eqnarray}\label{derextfields}
E^{ij} & = & -iE^i q^j\nonumber\\
B^{ij} & = & -iB^i q^j
\end{eqnarray}
are the Fourier transforms of their space derivatives
$\partial E^i(\vec x)/\partial x^j$
and $\partial B^i(\vec x)/\partial x^j$.

Thus, using Eq.~(\ref{Gammagen}) to evaluate the amplitude,
and then comparing with the form given in 
Eq.~(\ref{ampextfieldform}), we obtain for the
magnetic dipole and quadrupole moment matrices 
\begin{eqnarray}\label{dsubMmatrix}
\vec d_M & = & d_M\vec S\\
\label{QsubMmatrix}
x_M^{ij} & = & \frac{1}{3}Q_MI^{ij}\,,
\end{eqnarray}
where
\begin{eqnarray}\label{dsubM}
d_M & = & -\left(\frac{a_3(0)}{2m_V}\right)\\
\label{QsubM}
Q_M & = & -\frac{3}{2m_V}\left[\frac{b_1(0)}{m_V} - (b_2(0) + b_3(0))\right]\,.
\end{eqnarray}
It is easy to verify that the above result
for $\vec d_M$, for example, is the same that is obtained 
from the formula
\begin{equation}\label{dsubMexp}
\xi^{\prime\dagger}d_M^i \xi = \frac{1}{2m_V}
\varepsilon^{ijk}\left.\left[\frac{i\partial J^k(\vec q)}{\partial q^j}\right]\right|_
{\vec q = 0}\,,
\end{equation}
which follows by using the definition
\begin{equation}\label{dsubMdef}
\xi^{\prime\dagger}\vec d_M \xi
(2\pi)^3 2k^0\delta^{(3)}(\vec k - \vec k^\prime)
= \left. \langle V(k^\prime s^\prime)|\vec D_M|V(ks)\rangle
\right|_{|\vec k^{\prime}| = |\vec k|\rightarrow 0}\,,
\end{equation}
in terms of the magnetic dipole moment operator 
\begin{equation}\label{DsubMopdef}
\vec D_M = \frac{1}{2}\int d^3\! x\, \vec x\times \vec j^{(EM)}(\vec x)\,.
\end{equation}

We must mention that when $\vec J(\vec q)$ is expanded
in powers of $\vec q$ in Eq.~(\ref{ampextfield}), not all terms
that are generated can be written in the form of Eq.~(\ref{ampextfieldform}).
The amplitude contains additional terms such as, for example, 
\begin{equation}\label{nongaugeinvterms}
a_4 \xi^{\,\ast\prime j}\xi^i (A^iq^j + A^jq^i)\,,
\end{equation}
which do not fit in the form of Eq.~(\ref{ampextfieldform}).
However, all such terms in fact vanish once
the constraint of gauge invariance, which in the
static limit is simply
\begin{equation}\label{gaugeinvstatic}
\vec q\cdot\vec J(\vec q) = 0\,,
\end{equation}
is imposed upon them. While we do not discuss this any further
here, the implications of gauge invariance
are considered in all detail in Section \ref{sec:gaugeinv}.

%
%
\section{Implications of the discrete symmetries}\setcounter{equation}{0}
\label{sec:discretesymmetries}
In this section we deduce the constraints on the vertex
function, and in turn on the form 
factors, that are imposed by several general requirements
and/or symmetry principles.  These are: crossing symmetry,
the Hermiticity of the Langrangian and, when applicable,
the discrete space symmetries $C$, $P$ and $T$, or the
appropriate combinations of them.  We consider the
the diagonal and off-diagonal cases separately.
\subsection{Diagonal case: $V = V'$}
Let us consider the space-time symmetries first.
Consider, for example, the parity symmetry.  The
effect of its transformation can be summarized by
the statement
\begin{equation}\label{Ptransform}
\Gamma_{\alpha\alpha'\mu}(k,k')\stackrel{P}{\rightarrow}
\Gamma^P_{\alpha\alpha'\mu}(k,k')\,,
\end{equation}
where $\Gamma^P_{\alpha\alpha'\mu}$ is obtained from 
$\Gamma_{\alpha\alpha'\mu}$ by multiplying every quantity 
that appears in $\Gamma_{\alpha\alpha'\mu}$ by its parity 
phase $\eta_P$ tabulated in Table~\ref{table:transformations}.  
\begin{table}

\begin{center}
\begin{tabular}{llllll}\hline\hline 
  & $\eta_P$ & $\eta_T$ & $\eta_C$ & $\eta_{CP}$ & $\eta_{CPT}$\\ \hline
1 &   $+$    &  $+$     &    $+$   &    $+$      &    $+$      \\
$i$ &   $+$    &  $-$     &    $+$   &    $+$      &    $-$      \\
$\epsilon_{\alpha\beta\lambda\rho}$
  &   $-$    &  $-$     &    $+$   &    $-$      &    $+$      \\ \hline\hline
\end{tabular}
\end{center}
\caption{The transformation rules of different quantities
appearing in $\Gamma_{\alpha\alpha'\mu}$ under
the various discrete symmetries.\label{table:transformations}}
\end{table}
Using similar notation for the effect of the charge conjugation
and time reversal transformations, we obtain
\begin{eqnarray}\label{Ctransform}
\Gamma_{\alpha\alpha'\mu}(k,k') &\stackrel{C}{\rightarrow} &
-\Gamma^C_{\alpha'\alpha\mu}(-k',-k)\,, \\* 
\Gamma_{\alpha\alpha'\mu}(k,k') &\stackrel{T}{\rightarrow} &
-\Gamma^T_{\alpha'\alpha\mu}(-k,-k') \,.
\label{Ttransform}
\end{eqnarray}
In general, the $C$ transformation and the crossing
relation, each one separately, give a relation that expresses the amplitude for
the antiparticle process ($\overline V\rightarrow\overline V^\prime
+ \gamma$) in terms of $\Gamma_{\alpha\alpha^\prime\mu}$. 
Strictly speaking, the $C$ transformation rule written
above is the combined effect of the Charge-conjugation
transformation and the crossing relation.
As discussed below, 
in the case that $V$ is self-conjugate the crossing relation
gives by itself another independent constraint.

{}From these transformations we can deduce the effects
of combined operations, for example,
\begin{eqnarray}\label{CPtransform}
\Gamma_{\alpha\alpha'\mu}(k,k') & \stackrel{CP}{\rightarrow} &
-\Gamma^{CP}_{\alpha'\alpha\mu}(-k',-k)\,,\\
\nonumber\\
\label{CPTtransform}
\Gamma_{\alpha\alpha'\mu}(k,k') & \stackrel{CPT}{\rightarrow} &
\phantom{-}\Gamma^{CPT}_{\alpha'\alpha\mu}(k',k)\,.
\end{eqnarray}

If the Lagrangian is symmetric under any of the discrete
symetry transformations, then the arrow in the
corresponding relation in Eqs.\ (\ref{Ptransform}-\ref{CPTtransform})
should be replaced by the equality sign.  The resulting
equations produce constraints on the form factors.
For example, if $CP$ is a symmetry of the Lagrangian, then 
from the above it follows that the vertex function satisfies
\begin{equation}
\label{CPrelation}
\Gamma_{\alpha\alpha^\prime\mu}(k,k^\prime)
= -\Gamma^{CP}_{\alpha^\prime\alpha\mu}(-k^\prime,-k)\,,
\end{equation}
which implies that
\begin{equation}
a_1 = a_2 = a_4 = 0
\end{equation}
and
\begin{eqnarray}\label{CPconstraint}
b_1 & = & 0\nonumber\\
b_3 & = & -b_2\,.
\end{eqnarray}
{}From Eqs.\ (\ref{dsubE}) and (\ref{QsubM}), 
the relations in
Eq.\ (\ref{CPconstraint}) imply that the electric dipole moment
and the magnetic quadrupole moments are zero, which is
a familiar result.

On top of all these relations, which may or may not be
satisfied in a particular case, the following relation
follows from the fact that the Lagrangian is Hermitian:
\begin{equation}\label{Hcondition}
\Gamma_{\alpha\alpha'\mu}(k,k')
= \Gamma^\ast_{\alpha'\alpha\mu}(k',k)\,,
\end{equation} 
This relation, which is independent of whether the particle
is neutral or charged, or of the status of the discrete
symmetries, yields the following reality conditions on the form
factors:
\begin{eqnarray}\label{realityconds}
a'_1, a'_2, a_3, b_1 & = & \mbox{real}\nonumber\\
a_1, a_2, a_4, b'_1 & = & \mbox{imaginary}\nonumber\\
b_3 & = & b^\ast_2 \,,
\end{eqnarray}
which in particular imply that the electromagnetic moments,
as we have identified them, are real.
Notice also that once the Hermiticity condition
is imposed, the CPT transformation in Eq.~(\ref{CPTtransform})
becomes a trivial identity, as it should be.

The above discussion applies to both the charged
and neutral cases.  However, for the neutral case
there is an additional independent constraint if the particle
is not just neutral but also self-conjugate. 
If the particle is self-conjugate, then
the vertex obeys the crossing relation corresponding
to the exchange of the external $V$ lines.  
The resulting condition is simply 
\begin{eqnarray}\label{crossingrel}
\Gamma_{\alpha\alpha'\mu}(k,k') = \Gamma_{\alpha'\alpha\mu}(-k',-k) \,. 
\end{eqnarray}
%
%
%
Since all the form factors are now functions of $q^2$ only, they remain
invariant under the substitution $k\leftrightarrow -k'$.
Eq.~(\ref{crossingrel}) applied to Eq.~(\ref{Gammagen}) 
implies that the form factors $a_1^\prime, a'_2, a_3, b_1$ all vanish,
while $b_3 = -b_2$.  
Thus, the symmetry reduces $\Gamma^{(T)}_{\alpha\alpha'\mu}$
to the form
\begin{eqnarray}\label{GammagenneutBose}
\Gamma_{\alpha\alpha'\mu} & = &
a_1 q_\mu g_{\alpha\alpha'} +
a_2 q_\mu q_\alpha q_{\alpha'} +
a_4 (g_{\mu\alpha'}q_\alpha + g_{\mu\alpha}q_{\alpha'})\nonumber\\
& & \mbox{} + b^\prime_1\epsilon_{\alpha\alpha^\prime\mu\nu}Q^\nu + 
b_2\left(q_\alpha[Qq]_{\mu\alpha'} - 
q_{\alpha'}[Qq]_{\mu\alpha}\right) \,.
\end{eqnarray}

Notice that this result implies that
the charge form factor as well as
the dipole moment form factors are zero for all
$q^2$ and not just for $q^2 = 0$.  Moreover, it
implies in particular that a neutral (self-conjugate) vector
particle can have neither an electric nor a magnetic
dipole moment, in complete analogy with the result
for Majorana fermions \cite{Majo}.  Furthermore, since 
$a_1^\prime = a_2^\prime = a_3 = 0$
the electric quadrupole moment is zero,
and since $b_1 = 0$ and $b_3 = -b_2$ the magnetic quadrupole
moment is also zero.  In short, a self-conjugate
vector particle cannot have any static electromagnetic moments.
Nevertheless, it is interesting to note that even self-conjugate 
particles can have an electromagnetic vertex. 
However, as will
be shown in Section \ref{sec:gaugeinv}, gauge invariance
places further restrictions on the
other form factors $a_1$, $a_2$, $a_4$, $b_1^\prime$ and $b_2$,
and in the end only
$b_2$ and a linear combination of $a_2$ and $a_4$ can be non-zero.

In the Standard Model, the $Z$ falls in this class; it is
self-conjugate (which implies its charge neutrality). 
But in principle, there can exist vector particles that
are electrically neutral, but which have a non-zero
quantum number with respect to the charge of some
other (global or local) symmetry which is at the moment
unknown to us.  For such a particle, which is
neutral but not self-conjugate, the constraint
of Eq.~(\ref{crossingrel}) does not apply.  The distinction
between the neutral and self-conjugate vector particles
is analogous to the one between Dirac and Majorana neutrinos.
For particles that are not self-conjugate, the crossing
symmetry relations do not give us constraints on the
form factors, but instead allow us to relate the form
factors of the antiparticle (the conjugate particle)
to those of the particle.

One interesting case that appears as a special one is the off-shell 
coupling of three photons. If $C$-invariance holds, we can easily 
see that the vertex vanishes. This follows from Eqs.\ (\ref{Ctransform}) 
and (\ref{crossingrel}). From Table~\ref{table:transformations}, 
we see that $\eta_C$ is $+1$ for all quantities of interest, so 
the quantity $\Gamma^C$ appearing in Eq.\ (\ref{Ctransform}) is really 
equal to $\Gamma$. If $C$-invariance is valid, the two expressions 
in Eq.\ (\ref{Ctransform}) should be equal, which directly contradicts 
Eq.\ (\ref{crossingrel}) unless the entire vertex vanishes. This result 
is known as Furry's theorem\cite{furry}. 
However, note that this argument
shows that in order that the vertex vanishes,
$V$ and $V'$ need not be photons;
$C$-invariance and $V=V'$ are sufficient conditions.
In other words, the 
off-shell vertex $VV\gamma$ vanishes for any self-conjugate $V$ provided 
$C$-invariance holds.

On the other hand, since we know that $C$-invariance does not 
hold when the weak interactions are 
taken into account, the $VV\gamma$ vertex need not vanish 
in general. This will be discussed after introducing the constraints 
from gauge invariance in Sec.~\ref{sec:gaugeinv}.

\subsection{Off-diagonal case: $V \not = V'$}
In analogy with Eq.~(\ref{defGamma}) we define the vertex function
for the process
\begin{equation}\label{V'->VA}
V'(k') \rightarrow V(k) + \gamma(q)\,.
\end{equation}
by writing
\begin{equation}\label{definvGamma}
\langle V(k)|j_\mu^{EM}(0)|V'(k')\rangle
= \epsilon^{\prime\alpha'}(k')\epsilon^{\ast\alpha}(k)
\Gamma^{(V'\rightarrow V)}_{\alpha'\alpha\mu}(k',k)\,.
\end{equation}
Using a notation similar as before, the effect
of the various transformations on the vertex
function can be summarized by the following rules.
For $P$ and $T$ we have
\begin{equation}\label{Ptransformoff}
\Gamma_{\alpha\alpha'\mu}(k,k')\stackrel{P}{\rightarrow}
\delta^{\prime\ast}_P\delta_P
\Gamma^P_{\alpha\alpha'\mu}(k,k')\,,
\end{equation}
\begin{equation}\label{Ttransformoff}
\Gamma_{\alpha\alpha'\mu}(k,k')\stackrel{T}{\rightarrow}
-\delta^{\prime\ast}_T\delta_T
\Gamma^T_{\alpha'\alpha\mu}(-k,-k')\,,
\end{equation}
where $\delta_{P,T}$ and $\delta^{\prime}_{P,T}$ are the
phases that appear in the $P,T$ transformation rules of the 
$V,V'$ fields.  Under charge conjugation,
\begin{equation}\label{Ctransformoff}
\Gamma_{\alpha\alpha'\mu}(k,k')\stackrel{C}{\rightarrow}
-\delta^{\prime}_C\delta^\ast_C
\Gamma^{(V'\rightarrow V)}_{\alpha'\alpha\mu}(-k',-k)\,,
\end{equation}
while the Hermiticity condition becomes the statement
that
\begin{equation}\label{Hconditionoff}
\Gamma^{(V'\rightarrow V)}_{\alpha'\alpha\mu}(k',k)\,
= \Gamma^\ast_{\alpha\alpha'\mu}(k,k')
\end{equation}
The comment made after Eq.~(\ref{Ttransform}) applies here also. 
In particular, for the case of self-conjugate particles
the crossing relation gives the additional condition
\begin{equation}\label{Mconditionoff}
\Gamma^{(V'\rightarrow V)}_{\alpha'\alpha\mu}(k',k)
= \Gamma_{\alpha\alpha'\mu}(-k,-k')\,.
\end{equation}
Notice that taking $V = V^\prime$ as a 
particular case in Eq.~(\ref{Mconditionoff}), the condition expressed
in Eq.~(\ref{crossingrel}) is reproduced.

As an example of the kind of relation that we can deduce
from these results, suppose that $CP$ is conserved.
{}From Eqs.\ (\ref{Ptransformoff}) and (\ref{Ctransformoff}) it
then follows that
\begin{equation}\label{CPconditionoff}
\Gamma^{(V'\rightarrow V)}_{\alpha'\alpha\mu}(k',k)
= -\delta^{\prime\ast}_{CP}\delta_{CP}
\Gamma^P_{\alpha\alpha'\mu}(-k,-k')\,.
\end{equation}
Comparing this with
Eq.~(\ref{Hconditionoff}) we then obtain the condition
\begin{equation}
\label{CP+Hconditionoff}
\Gamma^\ast_{\alpha\alpha'\mu}(k,k')
= -\delta^{\prime\ast}_{CP}\delta_{CP}
\Gamma^P_{\alpha\alpha'\mu}(-k,-k')\,,
\end{equation}
which implies that $a_{1,2,3,4}$ and $a'_{1,2}$
have the same phase $e^{i\phi}$, while
$b_{1,2,3}$ and $b'_1$ all have the same phase
$ie^{i\phi}$, with the same $\phi$.  It is also easy
to verify that the results that were derived previously
for the diagonal case are reproduced here if we 
specialize these formulas to that case by setting $V'=V$.

The case of self-conjugate particles is more interesting.
As already commented, the new feature is that
Eq.~(\ref{Mconditionoff}) is valid independently of whether
the discrete symmetries are conserved or not. That condition 
together with the hermiticity condition of Eq.~(\ref{Hconditionoff}) give
\begin{equation}
\label{hermselfcong}
\Gamma_{\alpha\alpha'\mu}(k,k') =
\Gamma^\ast_{\alpha\alpha'\mu}(-k,-k')\,,
\end{equation}
which implies that all the form factors are
purely imaginary.  Now, if $CP$ is conserved,
we can combine this with what we concluded
after Eq.~(\ref{CP+Hconditionoff}).  There are two possibilities
depending on the relative sign of the $CP$ phases
of $V$ and $V'$, which we denoted by $\delta_{CP}$
and $\delta'_{CP}$ in Eq.~(\ref{CPconditionoff}).
\begin{description}
\item[Same $CP$ phase of $V'$ and $V$] In this case we have
\begin{eqnarray}
a_{1,2,3,4} = a'_{1,2} = 0\,.
\end{eqnarray}
Only the $b$ coefficients survive (and they are purely
imaginary), so that the transition
moments $d_M, Q_E$ are zero while $d_E, Q_M$ are non-zero.
\item[Oppposite $CP$ phases of $V'$ and $V$] In this case 
the opposite occurs:  the $b$ coefficients
are zero while the $a$ coefficients are non-zero
and purely imaginary. Accordingly, the transition
moments $d_E$, $Q_M$ are zero while $d_M$, $Q_E$ are allowed.
This is, once more, in complete analogy with the situation
for Majorana fermions\cite{Majo}.
\end{description}

The discussion above immediately brings the question
of whether the relative $CP$ phase of all self-conjugate
vector bosons is positive.  For the photon and the $Z$
it is true, but 
can one construct examples where it is not?
The following examples illustrate the various
possibilities that can arise.

Suppose that $V$ and $V^\prime$ have the following couplings
to a pair of neutrinos, $\nu_{Le,\mu}$
\begin{eqnarray}\label{model1}
L^\prime & = &
aV^{\prime\mu}\left[\overline\nu_{Le}\gamma_\mu\nu_{L\mu}
+ \overline\nu_{L\mu}\gamma_\mu\nu_{Le}\right]\nonumber\\
& & \mbox{} + ibV^{\mu}\left[\overline\nu_{Le}\gamma_\mu\nu_{L\mu}
- \overline\nu_{L\mu}\gamma_\mu\nu_{Le}\right]
\end{eqnarray}
It is clear from this that the $CP$ phase of $V$ and $V^\prime$
must be opposite in order to keep $L^\prime$ invariant
under $CP$.  
A generalization of this interaction
is to promote it to an $SU(2)$ gauge interaction by writing
\begin{equation}\label{model2}
L^\prime = -g\overline\eta_L\gamma_\mu\frac{\vec\tau}{2}\eta_L\cdot\vec V^\mu
\end{equation}
where $\eta_L$ stands for a doublet formed by $\nu_{Le,\mu}$.
In this model, under $CP$
\begin{equation}
V^{1,3}\rightarrow -V^{1,3}\,,
\end{equation}
just like the $Z$ and the photon, but
\begin{equation}
V^{2}\rightarrow V^{2}\,.
\end{equation}
Interactions of a similar type generally appear in models that
have been considered in various contexts, such as
grand unified theories, family or horizontal symmetries
and superstring inspired models.

In the examples considered in Eqs. (\ref{model1})
and (\ref{model2}), the vector bosons are self-conjugate.
On the other hand, if the theory is such that the individual
lepton numbers $L_e$ and $L_\mu$ are separately
conserved, then it is more convenient to work with
\begin{equation}
V_\mu\equiv \frac{1}{\sqrt{2}}(V_{\mu}^1 + iV_{\mu}^2)
\end{equation}
and its complex conjugate,
because they are the eigenstates of the conserved $L_{e,\mu}$ operators.
The couplings of $V$ to the neutrinos is of the form
$V_\mu \overline\nu_{L e}\gamma^\mu\nu_{L\mu} + h.c.$,
and in such a theory $V$ would carry both $L_e$ and $L_\mu$
quantum numbers.
While electrically neutral, $V$ is not self-conjugate.
If there is another such vector particle $V^\prime$,
one of them can decay radiatively into the other.

We would like to point out that, in the examples given in this
Section,
we have focused the attention on the relations for the
$a$ and $b$ coefficients, which are contained
in $\Gamma^{(T)}_{\alpha\alpha'\mu}$.
We have chosen that
only for illustrative purposes, motivated by the fact that
those are the form factors that contribute when the vector bosons
are on-shell, and therefore the ones that have a direct
physical interpretation in terms of the electromagnetic moments
of the particles.  
However, analogous relations can be similarly
derived for the $c$ and $d$ coefficients
contained in $\Gamma^{(L)}_{\alpha\alpha'\mu}$,
by applying the
conditions that the vertex function must satisfy,
such as Eq.~(\ref{CPrelation}) or (\ref{Hcondition}),
to Eq.\ (\ref{Gammagen2}).

%
%
\section{Gauge invariance}\setcounter{equation}{0}
\label{sec:gaugeinv}
Now we explore what are the constraints imposed by the
electromagnetic gauge invariance on the vertex function. 
The analysis depends on whether the
two gauge bosons $V$ and $V'$ are charged or neutral,
therefore we consider them separately.
\subsection{$V$, $V'$ are neutral}
The condition is expressed 
in Eq.~(\ref{gaugeinvstrong}), and
applying it to Eq.~(\ref{Gammatot}) we get the following 
relations:
\begin{eqnarray}\label{gaugeinvstrongconds}
a_1 q^2 + a'_1 Q\cdot q & = & 0\nonumber\\
a_2  q^2 + a'_2 Q\cdot q + 2a_4 & = & 0\nonumber\\
b'_1 & = & 0\nonumber\\
c_1q^2 + c'_1 Q\cdot q - c_4 + c_5 & = & 0\nonumber\\
c_2q^2 + c'_2 Q\cdot q + c_4  & = & 0\nonumber\\
c_3q^2 + c'_3 Q\cdot q - c_5 & = & 0\,.
\end{eqnarray}
The non-trivial ones of these relations are solved,
without introducing artificial singularities, by
writing
\begin{eqnarray}\label{gaugeinvrel}
a'_1 & = & q^2 a_0\nonumber\\
a_1 & = & -Q\cdot q a_0\nonumber\\
a_4 & = & - \frac{1}{2}(q^2 a_2 + Q\cdot q a^\prime_2)\nonumber\\
c_4 & = & -c_2 q^2 - c'_2 Q\cdot q\nonumber\\
c_5 & = & c_3 q^2 + c'_3 Q\cdot q\nonumber\\
c_1 + c_2 + c_3 & = & -c_0 Q\cdot q\nonumber\\
c'_1 + c'_2 + c'_3 & = & c_0  q^2 \,.
\end{eqnarray}
Therefore,
	\begin{eqnarray}\label{Gammagentrans}
\Gamma^{(T)}_{\alpha\alpha'\mu} & = & 
(q^2 Q_\mu - Q \cdot q q_\mu) a_0 g_{\alpha\alpha'}\nonumber\\
& & \mbox{}
+ a_2 \left( q_\mu q_\alpha q_{\alpha'} - {1\over2} q^2 
(g_{\mu\alpha'}q_\alpha + g_{\mu\alpha}q_{\alpha'}) \right)+
a'_2 \left( Q_\mu q_\alpha q_{\alpha'} - {1\over2} Q \cdot q
(g_{\mu\alpha'}q_\alpha + g_{\mu\alpha}q_{\alpha'}) \right) 
\nonumber\\
& & \mbox{} + a_3 (g_{\mu\alpha'}q_\alpha - g_{\mu\alpha}q_{\alpha'}) +
b_1\epsilon_{\alpha\alpha'\mu\nu}q^\nu + 
b_2 q_\alpha [Qq]_{\mu\alpha'} + 
b_3 q_{\alpha'} [Qq]_{\mu\alpha}\,.
\end{eqnarray}
and
\begin{eqnarray}\label{Gamma2trans}
\Gamma^{(L)}_{\alpha\alpha'\mu} & = & 
c_0 (q^2 Q_\mu - Q \cdot q q_\mu) k_\alpha k'_{\alpha'} +
c_2 k_\alpha (q_\mu q_{\alpha'} - q^2 g_{\mu\alpha'}) +
c'_2 k_\alpha (Q_\mu q_{\alpha'} - Q\cdot q g_{\mu\alpha'}) 
\nonumber\\* 
& & -
c_3 k'_{\alpha'} (q_\mu q_\alpha - q^2 g_{\mu\alpha}) -
c'_3 k'_{\alpha'} (Q_\mu q_\alpha - Q\cdot q g_{\mu\alpha}) +
d_1 k_\alpha [Qq]_{\mu\alpha'} + 
d_2 k'_{\alpha'} [Qq]_{\mu\alpha}\,.
\end{eqnarray}

In the most general case, the vertex cannot be simplified further. 
Simplification occurs in special cases as we discuss below.

\subsubsection{$V\neq V'$, all particles on-shell}
Physically, this represents a decay process where one vector
boson decays radiatively to another one. 
In this case we can reduce the vertex of Eq.\ (\ref{Gammagentrans})
further by the using the on-shell conditions for the photon,
	\begin{eqnarray}
\epsilon^{(\gamma)}\cdot q = 0 \,, \qquad q^2 = 0 \,.
\label{onshellphoton}
	\end{eqnarray}
This gives the form
\begin{eqnarray}\label{Gammaneutdecay}
\Gamma^{(T)}_{\alpha\alpha'\mu} & = & 
a'_2 \left( Q_\mu q_\alpha q_{\alpha'} - {1\over2} Q \cdot q
(g_{\mu\alpha'}q_\alpha + g_{\mu\alpha}q_{\alpha'}) \right) 
\nonumber\\*
& & + a_3 (g_{\mu\alpha'}q_\alpha - g_{\mu\alpha}q_{\alpha'}) +
b_1\epsilon_{\alpha\alpha'\mu\nu}q^\nu +
b_2 q_\alpha [Qq]_{\mu\alpha'} + 
b_3 q_{\alpha'} [Qq]_{\mu\alpha} 
\end{eqnarray}
while $\Gamma^{(L)}$ does not contribute.
 Notice that Eq.~(\ref{Gammaneutdecay}) involves a combination of dipole and
quadrupole moment terms only and, in particular, the term with 
the charge form factor $a'_1$ 
vanishes in this case.
The reason behind this is that in this configuration (all
particles on-shell)
the $a^\prime_1$ form factor is the matrix element of
the charge operator between the $V$ and $V^\prime$ states,
which is zero if the particles are different.

It is interesting to consider what occurs in this case
when the particle
$V'$ also is the photon, which corresponds to the
possibility of a vector boson decaying into two photons. In this
case, the vertex function should satisfy the additional constraint
	\begin{eqnarray}\label{2photonsymcond}
\Gamma_{\alpha\alpha'\mu} (k,k') = 
\Gamma_{\alpha\mu\alpha^\prime} (k,q) \,,
	\end{eqnarray}
which corresponds to the condition 
that the two photons in the final state obey Bose symmetry.
In addition, gauge invariance of the second photon requires
	\begin{eqnarray}\label{2photongaugeinv}
k'^{\alpha'} \Gamma_{\alpha\alpha'\mu} = 0 \,,
	\end{eqnarray}
but this is satisfied automatically if Eq.~(\ref{2photonsymcond})
is imposed upon (\ref{Gammaneutdecay}).
Eq.~(\ref{2photonsymcond}) translates to
\begin{eqnarray}\label{2photondecayBose}
\Gamma_{\alpha\alpha'\mu} = \left. \vphantom{1\over2}
\Gamma_{\alpha\mu\alpha^\prime}\right|_{Q\rightarrow\widetilde Q,
q\rightarrow\widetilde q}
	\end{eqnarray}
where
	\begin{eqnarray}
\widetilde Q &=& k+q = (Q + 3q)/2 \,, \\*
\widetilde q &=& k-q = (Q - q)/2 \,.
	\end{eqnarray}
After substituting Eq.~(\ref{Gammaneutdecay}) into 
Eq.\ (\ref{2photondecayBose}),
the right-hand side of Eq.~(\ref{2photondecayBose}) can be
reduced with the help of Eq.~(\ref{epsQ})
and the relations in Eqs.\ (\ref{polvecrelations}) 
plus additional relations such as
	\begin{eqnarray}
[\widetilde Q \widetilde q]_{\mu\nu} & = & - [Qq]_{\mu\nu} \,.
	\end{eqnarray}
It then follows easily from Eq.~(\ref{2photondecayBose}) 
that the vertex in fact vanishes. 
In other words, a spin-1 particle cannot decay to two photons. This is
the well-known Yang theorem~\cite{Yang}.

\subsubsection{$V=V'$, both on shell}
In this case the photon is not on-shell ($q^2 \not = 0$)
but, since both $V$ lines are on shell, 
$Q\cdot q = k^2 - k'^2 = 0$.  The vertex function
then reduces to
\begin{eqnarray}\label{Gammagenneutdiag}
\Gamma_{\alpha\alpha'\mu} & = & 
q^2 Q_\mu a_0 g_{\alpha\alpha'} + a_2 \left( q_\mu q_\alpha q_{\alpha'} - {1\over2} q^2 
(g_{\mu\alpha'}q_\alpha + g_{\mu\alpha}q_{\alpha'}) \right)\nonumber\\
& & \mbox{} + a'_2 Q_\mu q_\alpha q_{\alpha'} +
a_3 (g_{\mu\alpha'}q_\alpha - g_{\mu\alpha}q_{\alpha'})\nonumber\\
& & \mbox{} +
b_1\epsilon_{\alpha\alpha'\mu\nu}q^\nu + 
b_2 q_\alpha [Qq]_{\mu\alpha'} + 
b_3 q_{\alpha'} [Qq]_{\mu\alpha}\,.
\end{eqnarray}
As already discussed in Section \ref{sec:discretesymmetries},
if the particle is self-conjugate then
the vertex obeys the crossing relation corresponding
to the exchange of the external $V$ lines.  In this case,
Eq.~(\ref{crossingrel}) applied to Eq.~(\ref{Gammagenneutdiag}) 
implies that the form factors $a_0, a'_2, a_3, b_1$ all vanish,
while $b_3 = -b_2$, using the fact that in this case all form factors 
must be functions of $q^2$ only. 
Thus Eq.~(\ref{Gammagenneutdiag}) reduces
to the form
	\begin{eqnarray}\label{GammagenneutBosegaugeinv}
\Gamma_{\alpha\alpha'\mu} & = &
a_2 \left\{ q_\mu q_\alpha q_{\alpha'} - {1\over 2} q^2 
(g_{\mu\alpha'}q_\alpha + g_{\mu\alpha}q_{\alpha'}) \right\}
+ 
b_2\left(q_\alpha[Qq]_{\mu\alpha'} - \vphantom{1\over2}
q_{\alpha'}[Qq]_{\mu\alpha}\right) \,.
\end{eqnarray}
Therefore, while a self-conjugate particle cannot 
cannot have any static electromagnetic moments,
in the most general case it can have an
electromagnetic vertex characterized by
$a_2$ and $b_2$ as above.

\subsubsection{$V=V'$, all particles off-shell}
	This corresponds, for example, to the off-shell $ZZ\gamma$ 
vertex in the standard model or beyond. We should now use the vertex 
given in Eqs.\ (\ref{Gammagentrans}) and (\ref{Gamma2trans}), 
remembering that the crossing symmetry of 
Eq.\ (\ref{crossingrel}) applies here provided the particle $V$ is 
self-conjugate. The difference with the previous case is that now 
the form factors are functions of the Lorentz invariants $k^2$,
$k'^2$ and $q^2$, so the form factors $a_0$, $a'_2$, $a_3$, $b_1$ 
do not vanish, but rather satisfy the constraints
	\begin{eqnarray}
a_0 (k^2,k'^2,q^2) = - a_0 (k'^2,k^2,q^2)
\label{offshella0}
	\end{eqnarray}
and similarly for the other three. Similarly, the condition $b_3=-b_2$
obtained for the previous case should be replaced in this case by
	\begin{eqnarray}
b_2 (k^2,k'^2,q^2) = - b_3 (k'^2,k^2,q^2) \,.
\label{offshellb2b3}
	\end{eqnarray}
Moreover, the terms in $\Gamma^{(L)}$ also should be present in this
case, and the form factors in this part should satisfy the relations
	\begin{eqnarray}
c_0 (k^2,k'^2,q^2) &=& - c_0 (k'^2,k^2,q^2) \,, \nonumber\\*
c_2 (k^2,k'^2,q^2) &=& - c_3 (k'^2,k^2,q^2) \,, \nonumber\\*
c'_2 (k^2,k'^2,q^2) &=&  c'_3 (k'^2,k^2,q^2) \,, \nonumber\\*
d_1 (k^2,k'^2,q^2) &=& d_2 (k'^2,k^2,q^2) \,. \label{dcross}
	\end{eqnarray}

\subsubsection{$V'=\gamma$, both photons on-shell}
When $V'$ corresponds also to the photon, 
the vertex function, given by the sum 
of Eqs.\ (\ref{Gammagentrans}) 
and (\ref{Gamma2trans}), satisfies the additional constraint
	\begin{eqnarray}\label{transvadd}
k'^{\alpha'} \Gamma_{\alpha\alpha'\mu} = 0 \,.
\label{k'Gam=0}
	\end{eqnarray}
Moreover, since two of the bosons are photons, the vertex 
must satisfy the symmetry condition given 
in Eq.\ (\ref{2photonsymcond}).  Let us consider
Eq.\ (\ref{transvadd}) first. 
The contraction on the left hand side produces six different 
kind of tensor structures. Accordingly, we get six equations, but it
turns out that
only four of these are independent. Out of those four,
one connects the $b$ and 
the $d$ type coefficients by
	\begin{eqnarray}
{1\over 2} b_1 + b_3 k'\cdot q + d_2 k'^2 &=& 0 \,,
\label{bdcond1}
	\end{eqnarray}
which can be trivially solved for $b_1$. 
The other three can be written as
	\begin{eqnarray}
(c_3 + c'_3) k'^2 &=& (2a_0 + a_2 + a'_2) k' \cdot q \nonumber\\* 
(c_3 - c'_3) k'^2 &=& 2 k\cdot q a_0 + {1\over 2} (k^2-k'^2) a_2 
+ {1\over 2} q^2 a'_2 - a_3 \nonumber\\* 
c_2 + c'_2 &=& 2a_0 + 2c_0 k'^2 \,.
\label{acrelns1}
	\end{eqnarray}
These can be solved, without introducing artificial singularities, 
by writing
\begin{eqnarray}
\label{Vprimeqgam}
c_2 &=& a_0 + c_0 k'^2 + C_2 \nonumber\\*
c'_2 &=& a_0 + c_0 k'^2 - C_2 \nonumber\\*
c_3 &=& k' \cdot q A_3 + C_3 \nonumber\\*
c'_3 &=& k' \cdot q A_3 - C_3 \nonumber\\*
2a_0 + a_2 + a'_2 &=& 2 k'^2 A_3 \nonumber\\*
a_3 &=& 2 k\cdot q a_0 + {1\over 2} (k^2-k'^2) a_2 
+ {1\over 2} q^2 a'_2 - 2C_3 k'^2 \,,
\end{eqnarray}
so that $a_0$, $a_2$, $c_0$, $C_2$, $C_3$ and $A_3$ can be taken as the 
independent form factors in the $a$-$c$ sector. 
Using these form factors we can then write
	\begin{eqnarray}
\Gamma_{\alpha\alpha'\mu} &=& 2a_0 \left\{ 
g_{\alpha\alpha'} (k\cdot q k'_\mu - k' \cdot q k_\mu) +
q_{\alpha'} (k_\mu k_\alpha - Q_\mu q_\alpha) + 
g_{\mu\alpha'} (Q\cdot q q_\alpha - k\cdot q k_\alpha)
\right\}
\nonumber\\
& & +
2 (a_2 q_\alpha + C_2 k_\alpha) 
(g_{\mu\alpha'} k'\cdot q - k'_\mu q_{\alpha'}) 
\nonumber\\ 
&& + 2A_3 \left\{ k\cdot q g_{\mu\alpha} (k'\cdot q k'_{\alpha'}
- k'^2 q_{\alpha'}) -  k'\cdot q k'^2 
g_{\mu\alpha'} q_\alpha +  q_\alpha (k'^2 Q_\mu q_{\alpha'} - 
k' \cdot q k_\mu k'_{\alpha'})
\right\}
\nonumber\\ 
&& + c_0 k_\alpha \left\{ k'_{\alpha'} (q^2 Q_\mu - Q \cdot q q_\mu) +
2k'^2 (k_\mu q_{\alpha'} - k\cdot q g_{\mu\alpha'}) \right\} 
\nonumber\\* 
& & + 
2C_3 \left\{ k'_{\alpha'} (k'_\mu q_\alpha - k'\cdot q g_{\mu\alpha}) 
- k'^2 (g_{\mu\alpha'}q_\alpha - g_{\mu\alpha}q_{\alpha'})  \right\} 
\nonumber\\ 
& & +
b_2 q_\alpha [Qq]_{\mu\alpha'} + 
b_3 \left\{ q_{\alpha'} [Qq]_{\mu\alpha} -2 k'\cdot q
\epsilon_{\alpha\alpha'\mu\nu}q^\nu \right\}
\nonumber\\ && + 
d_1 k_\alpha [Qq]_{\mu\alpha'} + 
d_2 \left\{ k'_{\alpha'} [Qq]_{\mu\alpha} - 
2 k'^2 \epsilon_{\alpha\alpha'\mu\nu}q^\nu \right\} \,.
	\end{eqnarray}
Up to this point we have not used the fact 
that the two particles in the final state are both photons.
Therefore, the above expression is appropriate even in the case
that $V'$ is not the photon, provided that it couples
in the Lagrangian to a conserved current.
On the other hand, if $V'$ is indeed the photon also, 
then the condition in Eq.\ (\ref{2photonsymcond}) applies.  
In the case that we are considering, in which both photons are on-shell,
this condition implies that all form
factors, except $C_2$ and $d_1$, vanish. Thus, the general
form of the vertex function in this case is
\begin{eqnarray}
\Gamma_{\alpha\alpha'\mu} &=& 
2 C_2 k_\alpha  
(g_{\mu\alpha'} k'\cdot q - k'_\mu q_{\alpha'}) 
+ 2d_1 k_\alpha [k'q]_{\mu\alpha'} \,.
\end{eqnarray}

Notice that if the particle $V$ is on-shell, the vertex vanishes 
upon contracting with the polarization 
vector of the particle $V$, reproducing once more Yang's theorem.
On the other hand, for an off-shell $V$, the amplitude is not zero
as long as the $V$ line in the corresponding Feynman 
diagram is not attached to a conserved current.  This is
the situation if, for example, $V$ is attached to a neutrino
current $\overline\nu'_L\gamma_\mu\nu_L$ and at least one 
of the neutrinos has a non-zero mass.  Carrying this argument
a little further, it explains why the amplitude for the
process $\nu'\rightarrow\nu\gamma\gamma$ is proportional to
the neutrino masses in the local limit of the $W$-boson propagator
(leading order in $M^2_W$), which is a well known general
result\cite{gellmantheorem} and is also corroborated by explicit calculations\cite{twophotdecay}.

\subsubsection{Off-shell $V$, $V'$ which couple to 
conserved currents}
If both $V$ and $V'$ couple to conserve currents like the photon,
then the vertex function, given by the sum
of Eqs.\ (\ref{Gammagentrans}) 
and (\ref{Gamma2trans}), satisfies the condition
	\begin{eqnarray}
k^\alpha \Gamma_{\alpha\alpha'\mu} = 0 
	\end{eqnarray}
in addition to the condition given in Eq.\ (\ref{k'Gam=0}). 
This now gives some extra constraints on the form factors. For the 
$b$-type and $d$-type coefficients, one obtains 
	\begin{eqnarray}
\label{bdcond2}
-{1\over 2} b_1 + b_2 k\cdot q + d_1 k^2 &=& 0 
	\end{eqnarray}
in addition to Eq.\ (\ref{bdcond1}). 
These two equations imply that, in the $b$-$d$ sector, there are
three independent form factors. To express the form factors appearing 
in these equations in terms of three independent ones without 
introducing any artificial kinematic singularity, we first add 
the two equations to eliminate $b_1$, which gives
	\begin{eqnarray}
k^2 (d_1+b_2) + k'^2 (d_2 - b_3) + k\cdot k' (b_3-b_2) &=& 0 \,.
\label{bdrelns}
	\end{eqnarray}
This equation can be written as $\bf B \cdot K=0$, where
	\begin{eqnarray}
{\bf B} \equiv \left( \begin{array}{c}  
d_1+b_2 \\ d_2 - b_3 \\ b_3-b_2
\end{array} \right) \,, \qquad 
{\bf K} \equiv \left( \begin{array}{c}  
k^2 \\ k'^2 \\ k \cdot k' 
\end{array} \right) \,.
	\end{eqnarray}
We can then try to find two column matrices $\bf K_1$ and $\bf K_2$ 
which are both orthogonal to $\bf K$, and then the most general form 
for $\bf B$ would be $B_1 {\bf K}_1+ B_2 {\bf K}_2$, in terms of 
two new form factors $B_1$ and $B_2$. Choosing
	\begin{eqnarray}
\label{K1K2}
{\bf K}_2 \equiv \left( \begin{array}{c}  
-k'^2 \\ k^2 \\ 0 
\end{array} \right) \,, \qquad 
{\bf K}_1 \equiv 2\left( \begin{array}{c}  
k^2 k \cdot k' \\ k'^2 k \cdot k' \\ -(k^4+k'^4) 
\end{array} \right) \,, 
	\end{eqnarray}
we thus get\footnote{If we are working in the kinematic
region that includes the point $k^2=k'^2=0$, then
we cannot use ${\bf K}_1$ and ${\bf
K}_2$ as defined in Eq.\ (\ref{K1K2}), since they become null vectors
in this case. The analysis below has to be
modified accordingly.}
	\begin{eqnarray}
d_1+b_2 & = & 2 k^2 \, k\cdot k' B_1 - k'^2 B_2 \,, \nonumber\\*
d_2-b_3 & = & 2 k'^2 \, k\cdot k' B_1 + k^2 B_2 \,, \nonumber\\*
b_3-b_2 & = & - 2(k^4+k'^4) B_1 \,.
	\end{eqnarray}
The other form factor can be chosen by defining
	\begin{eqnarray}
b_3 + b_2 & \equiv & 2B_0 \,. 
	\end{eqnarray}
Together with Eqs.\ (\ref{bdcond1}) and (\ref{bdcond2}), 
these definitions give
	\begin{eqnarray}
b_2 &=& B_0 + (k^4+k'^4) B_1 \,,\nonumber \\ 
b_3 &=& B_0 - (k^4+k'^4) B_1 \,, \nonumber\\ 
d_1 &=& -B_0 + (2k^2 \, k\cdot k' - k^4 - k'^4) B_1 - k'^2 B_2 \,,\nonumber\\
d_2 &=& B_0 + (2k^2 \, k\cdot k' - k^4 - k'^4) B_1 + k^2 B_2 \,,\nonumber\\
b_1 &=& -2k\cdot k' B_0 + 2k\cdot k' (k^4-k'^4) B_1 - k^2 k'^2 B_2 \,.
	\end{eqnarray}

For the $a$ and $c$-type coefficients, in addition to the relations 
given in Eq.\ (\ref{acrelns1}), we get two more independent 
conditions, which are:
	\begin{eqnarray}
(c_2 - c'_2) k^2 &=& (2a_0 - a_2 + a'_2) k\cdot q \nonumber\\*
c_3 - c'_3 &=& -2a_0 - 2c_0 k^2 \,.
\label{acrelns2} 
	\end{eqnarray}
Thus, for example, the form factor $C_3$ introduced in
Eq.\ (\ref{Vprimeqgam}) is identified in terms of $a_0$ and $c_0$ here, and 
similarly
we can eliminate $C_2$ by introducing a new form factor $A_2$
through the relations 
	\begin{eqnarray}
c_2 - c'_2 &=& 2 k\cdot q A_2 \nonumber\\*
2a_0 - a_2 + a'_2 &=& 2k^2 A_2 \,.
	\end{eqnarray}
So finally four form factors remain independent, which can be taken 
as $a_0$, $c_0$, $A_2$ and $A_3$. The other ones, in terms of these
four independent ones, are identified by these relations:
	\begin{eqnarray}
a_0 &=& - A_0 \,, \nonumber\\* 
c_0 &=& A_1 \,, \nonumber\\ 
a_2 &=& k'^2 A_3 - k^2 A_2 \,, \nonumber\\ 
a'_2 &=& 2A_0 + k'^2 A_3 + k^2 A_2 \,, \nonumber\\ 
a_3 &=& 2k^2k'^2 A_1 - (k^2+k'^2) A_0 + k'^2 \, k\cdot q A_3 
	- k^2 \, k'\cdot q A_2 \,, \nonumber\\
c_2 &=& -A_0 + k \cdot q A_2 + k'^2 A_1  \,, \nonumber\\*
c'_2 &=& -A_0 -k \cdot q A_2 + k'^2 A_1  \,, \nonumber\\*
c_3 &=& A_0 + k'\cdot q A_3 - k^2 A_1  \,,\nonumber\\* 
c'_3 &=& -A_0 + k'\cdot q A_3 + k^2 A_1  \,.
	\end{eqnarray}
Putting these into Eqs.\ (\ref{Gammagentrans}) and (\ref{Gamma2trans}), we 
thus obtain the most general form of the vertex for this case:
	\begin{eqnarray}\label{offshellvertex}
\Gamma_{\alpha\alpha'\mu} & = & 
2A_0 \left\{ g_{\alpha\alpha'} (k'\cdot q k_\mu - k\cdot q k'_\mu) 
+ g_{\mu\alpha} (k\cdot k' q_{\alpha'} - k'\cdot q k_{\alpha'}) 
+ g_{\mu\alpha'} (k\cdot q k'_\alpha - k\cdot k' q_\alpha) \right.
\nonumber\\*  && \qquad + 
\left. k_{\alpha'} k'_\mu q_\alpha - k_\mu k'_\alpha q_{\alpha'}    
\right\} \nonumber\\* 
&& + 2A_1 \left\{ k'^2 [ g_{\mu\alpha'} 
\left( k^2 q_\alpha - k \cdot q k_\alpha \right) + k_\alpha k_{\alpha'} 
k_\mu ]
 - k^2 [ g_{\mu\alpha} 
\left( k'^2 q_{\alpha'} - k' \cdot q k'_{\alpha'} \right) 
- k'_\alpha k'_{\alpha'} k'_\mu ] \right. \nonumber\\* && \qquad
\left. - k \cdot k' (k_\mu+k'_\mu) k_\alpha k'_{\alpha'} \right\} 
\nonumber\\* 
&& + 2A_2 \left( k^2 q_\alpha - k \cdot q k_\alpha \right)
\left( k'_\mu q_{\alpha'} - k'\cdot q g_{\mu\alpha'} \right) \nonumber\\* 
&& + 2A_3 \left( k'^2 q_{\alpha'} - k' \cdot q k'_{\alpha'} \right)
\left( k_\mu q_\alpha - k\cdot q g_{\mu\alpha} \right) \nonumber\\* 
&& - B_0 \left\{ 2 k\cdot k' \varepsilon_{\alpha\alpha'\mu\nu} q^\nu 
+ k'_\alpha [Qq]_{\mu\alpha'} - k_{\alpha'} [Qq]_{\mu\alpha} \right\} 
\nonumber\\* 
&& + 2 k\cdot k' B_1 \left\{ 
(k^4-k'^4) \varepsilon_{\alpha\alpha'\mu\nu} q^\nu + 
k^2  k_\alpha [Qq]_{\mu\alpha'} - 
 k'^2 k'_{\alpha'} [Qq]_{\mu\alpha}
\right\} \nonumber\\* 
&& \qquad
+ (k^4+k'^4) B_1 \left\{  k_{\alpha'} [Qq]_{\mu\alpha} 
- k'_\alpha [Qq]_{\mu\alpha'}  \right\} 
\nonumber \\* 
&& + B_2 \left\{ k^2 k'^2 \varepsilon_{\alpha\alpha'\mu\nu} q^\nu - 
k'^2 k_\alpha [Qq]_{\mu\alpha'} + k^2 k'_{\alpha'} [Qq]_{\mu\alpha}
\right\}  \,.
\label{3consv}
	\end{eqnarray}

For the particular case $V=V'$, there are additional
restrictions which can 
be easily obtained from the conditions on the form factors 
expressed in Eqs.\ (\ref{offshella0}) through (\ref{dcross}).
These conditions imply
the following relations 
for the new form factors appearing in Eq.\ (\ref{3consv}), 
\begin{eqnarray}
A_{0,1}(k^2,k^{\prime 2},q^2) & = & -A_{0,1}(k^{\prime 2},k^2,q^2)\nonumber\\
A_{2,3}(k^2,k^{\prime 2},q^2) & = & -A_{3,2}(k^{\prime 2},k^2,q^2)\nonumber\\
B_{0}(k^2,k^{\prime 2},q^2) & = & -B_{0}(k^{\prime 2},k^2,q^2)\nonumber\\
B_{1,2}(k^2,k^{\prime 2},q^2) & = & -B_{1,2}(k^{\prime 2},k^2,q^2)\,,
\end{eqnarray}
which we represent schematically 
by saying that, under the interchange 
$k^2\leftrightarrow k^{\prime 2}$,
	\begin{eqnarray}
& A_0 \to - A_0, \quad A_1 \to -A_1, \quad A_2 \leftrightarrow -A_3
\nonumber \\*
& B_0 \to -B_0, \quad B_1 \leftrightarrow -B_1,  \quad 
B_2 \leftrightarrow -B_2  \,.
\label{3crossing}
	\end{eqnarray}

Finally, if $V=V'$ is the photon itself then Eq.\ (\ref{offshellvertex})
gives the
3-photon vertex.  However, in this case
the vertex must satisfy also the
condition expressed in Eq.\ (\ref{2photonsymcond})
and the additional crossing relation
\begin{equation}
\label{addcrosscond}
\Gamma_{\alpha\alpha^\prime\mu}(k,k^\prime) = 
\Gamma_{\mu\alpha^\prime\alpha}(-q,k^\prime)\,.
\end{equation}
Applying them to
Eq.\ (\ref{offshellvertex}), it follows that
the relations given in Eq.\ (\ref{3crossing}) should in fact
be valid if any two 
external momenta are interchanged. 
Using these extra crossing relations we then obtain
	\begin{eqnarray}
A_1 = A_2 = A_3 = B_1 = B_2 = 0 \,.
	\end{eqnarray}
Thus, in general,  the off-shell 3-photon vertex
does not vanish and it is characterized by the two form factors
$A_0$ and $B_0$.
It can 
be easily seen that the effective interaction in this case can be
rewritten as
	\begin{eqnarray}
2A_0 F^\lambda{}_\nu (k) F^\nu{}_\rho (k') F^\rho{}_\lambda (q)
- 2B_0 \tilde F^\lambda{}_\nu (k) \tilde F^\nu{}_\rho (k') 
\tilde F^\rho{}_\lambda (q)\,,
	\end{eqnarray}
where
	\begin{eqnarray}
F_{\lambda\nu} (k) & = & -i (k_\lambda A_\nu(k) - 
k_\nu A_\lambda(k)) \,, \nonumber\\* 
\tilde F_{\lambda\nu} (k) & = & {1\over 2} 
\varepsilon_{\lambda\nu\rho\sigma} 
F^{\rho\sigma} (k) \,.
	\end{eqnarray}
One may wonder, why not also interactions of the types $FF\tilde F$ and 
$\tilde F\tilde FF$? It can be shown that the first of these has 
precisely the form of the interaction with three factors of $\tilde F$, 
whereas the second is equal to the interaction with three factors of 
$F$.

\subsection{$V, V^\prime$ charged}
When $V$ and $V^\prime$ are charged, the condition on the
vertex function due to
gauge invariance is expressed in Eq.~(\ref{gaugeinvweak}).
Thus, what we get are conditions on the form factors
evaluated for  $k^2 = m_V^2$
and $k^{\prime 2} = m_{V^\prime}^2$.  Since the implications
depend on whether $m_V = m_{V^\prime}$, we
consider several cases separately.

\subsubsection{$V \neq V^\prime$, both on-shell}
Using Eq.~(\ref{Gammagen}) in Eq.~(\ref{gaugeinvweak}),
we get the relations 
\begin{eqnarray}\label{gaugeinvstrongcondscharged}
a_1 q^2 + a'_1 Q\cdot q & = & 0\nonumber\\
a_2  q^2 + a'_2 Q\cdot q + 2a_4 & = & 0\nonumber\\
b'_1 & = & 0 \,,
\end{eqnarray}
remembering that these are the form factors evaluated
with $k,k^\prime$ on shell.
Since in this case
$k^2 \not = k^{\prime 2}$, then $Q\cdot q \not = 0$.
Thus, we then have for $a_{1,4}$ and $a_1^\prime$
a set of relations analogous
to Eq.~(\ref{gaugeinvrel}), and we finally arrive at an expression
for the vertex function 
that is identical in form to $\Gamma^{(T)}_{\alpha\alpha'\mu}$
given in Eq.~(\ref{Gammagentrans}).

If the photon is also on-shell, corrresponding
to the decay process $V\rightarrow V^\prime\gamma$,
then vertex function reduces to exactly the same
form as given in Eq.~(\ref{Gammaneutdecay}).  The comments
made after Eq.~(\ref{Gammaneutdecay}) are applicable
in this case also.

\subsubsection{$V = V^\prime$, both on-shell}
In this case $k^2 = k^{\prime 2}$,
and therefore $Q\cdot q = 0$.  
Instead of Eq.~(\ref{gaugeinvstrongcondscharged})
we now have, for $k, k^\prime$ on-shell,
\begin{eqnarray}\label{gaugeinvweakconds}
a_1 = b_1^\prime & = & 0\nonumber\\
a_2 q^2 + 2a_4 & = & 0\,,
\end{eqnarray}
and
\begin{eqnarray}\label{Gammagentranschargdiag}
\Gamma_{\alpha\alpha'\mu} 
& = & 
a^\prime_1 Q_\mu g_{\alpha\alpha'}
 +
a'_2 Q_\mu q_\alpha q_{\alpha'} + a_3 (g_{\mu\alpha'}q_\alpha 
 - g_{\mu\alpha}q_{\alpha'}) \nonumber\\
& & \mbox{} 
+ a_4 \left( g_{\mu\alpha'}q_\alpha + g_{\mu\alpha}q_{\alpha'}
-2\frac{q_\mu q_\alpha q_{\alpha'}}{q^2}\right)
+ 
b_1\epsilon_{\alpha\alpha'\mu\nu}q^\nu + 
b_2 q_\alpha [Qq]_{\mu\alpha'} + 
b_3 q_{\alpha'} [Qq]_{\mu\alpha}\,.
\end{eqnarray}
The main difference between this case and the previous one
is that here the $a_1^\prime$ term does not vanish at $q^2 = 0$.
This is how it should be
since, as shown in Eq.~(\ref{Echarge}), in this limit the $a_1^\prime$ 
term corresponds to the
electric charge of the particle.
Notice also that, in spite of appearances, the $a_4$ term is well 
defined for on-shell photons because the apparently troublesome
term $q^\mu/q^2$ vanishes when it is contracted with the photon
polarization vector.  In fact, that term does not contribute whenever
the photon line is connected to a conserved current, such as one
generated by a pair of fermions on-shell.  However, in a more
complicated diagram in which the photon line connects to an off-shell
charged particle propagator, the contribution from the $q^\mu$ term 
in not zero and must be retained.  The apparent singularity
at $q^2 = 0$ is eliminated by the integration over the
internal loop momenta.

It is useful at this point to compare our expression in 
Eq.\ (\ref{Gammagentranschargdiag}) with the expression given
by 
Hagiwara, Peccei, Zeppenfeld and Hikasa\cite{HPHZ87}, which is much 
in use by other authors working in the field and therefore serves
as a good reference point.
Eq. (2.4) of Ref.\ \cite{HPHZ87} gives the vertex function
$\Gamma_{\alpha\beta\mu}$
for the process
\begin{equation}
\gamma_\mu(P)\rightarrow W_\alpha(q) + {\overline W}_\beta({\overline q})\,.
\end{equation}
The vertex function for the process 
${\overline W}_\alpha(q)\rightarrow \gamma_\mu(P) + 
{\overline W}_\beta({\overline q})$ is then given by
making the substitution $P\rightarrow -P, Q\rightarrow -Q$
in their Eq. (2.4).  Finally, by setting $\beta\rightarrow\alpha'$
and a making trivial relabeling of the momentum vectors 
in the resulting expression(which in the end amounts to
simply set $P\rightarrow q$), we 
obtain the vertex function for the process
${\overline W}_\alpha(k)\rightarrow \gamma_\mu(q) + 
{\overline W}_\beta(k')$, which corresponds to the
process we are considering with the identification
$V = V' = {\overline W}$.  Thus we obtain that the expression
that must be compared with our Eq.\ (\ref{Gammagentranschargdiag})
is
\begin{eqnarray}\label{HPZH}
\Gamma^{\rm (HPZH)}_{\alpha\alpha'\mu} & = & 
-f_1 Q_\mu g_{\alpha\alpha'}
+ {f_2 \over M_W^2} Q_\mu q_\alpha q_{\alpha'}
- f_3 (q_\alpha g_{\mu\alpha'} - q_{\alpha'}g_{\mu\alpha}) 
-if_4\left(q_\alpha g_{\mu\alpha'} + q_{\alpha'}g_{\mu\alpha}
-2\frac{q_\mu q_\alpha q_{\alpha'}}{q^2}\right)\nonumber\\
& & \mbox{}
-if_5\left(\epsilon_{\mu\alpha\alpha'\rho}Q^\rho - 
\frac{q^\mu}{q^2}[qQ]_{\alpha\alpha'}
\right)
+f_6 \epsilon_{\mu\alpha\alpha'\rho}q^\rho + 
\frac{f_7}{M_W^2}Q_\mu[qQ]_{\alpha\alpha'}\,.
\end{eqnarray}
In Eq.\ (\ref{HPZH}) we have included a term proportional
to $q^\mu$ in the factors of the $f_{4,5}$ terms. Such
terms were implicit in Ref.\ \cite{HPHZ87} but they were 
omitted under the assumption that the photon 
coupled to a conserved fermion current and therefore do
not contribute to the amplitude, as it is the case
for the process $e^{-}e^{+}\rightarrow W{\overline W}$
considered there.  We have restored them here 
for the purpose of our comparisson since we have not made
that assumption in the corresponding Eq.\ (\ref{Gammagentranschargdiag}).

Simple inspection of 
Eqs.\ (\ref{Gammagentranschargdiag}) and (\ref{HPZH}), 
reveal the following direct correspondence
between the form factors:
\begin{eqnarray}
a'_1 & = & -f_1\nonumber\\
a_4 & = & -if_4\nonumber\\
a'_2 & = & \frac{f_2}{M_W^2}\nonumber\\
a_3 & = & -f_3\,.
\end{eqnarray}
The correspondence between the remaining ones is not immediately
obvious, but follows straightforwardly upon using
the identities
\begin{eqnarray}\label{epsq2}
q^2\varepsilon_{\alpha\alpha'\mu\nu}Q^\nu - q_\mu [qQ]_{\alpha\alpha'}  & = & 
- q_{\alpha'} [Qq]_{\mu\alpha}
+ q_\alpha [Qq]_{\mu\alpha'}\\
\label{epsQ2}
Q_\mu [Qq]_{\alpha\alpha'} & = &
Q^2\varepsilon_{\alpha\alpha'\mu\nu}q^\nu
- q_{\alpha'} [Qq]_{\mu\alpha} 
- q_\alpha [Qq]_{\mu\alpha'} \,, 
\end{eqnarray}
which follow from Eqs.\ (\ref{epsq}) and (\ref{epsQ}) by specializing
them to the present situation ($V = V'$, both on-shell).
In this way we then obtain
\begin{eqnarray}
b_1 & = & f_6 - \frac{Q^2}{M^2_W}f_7\nonumber\\
b_2 & = & -i\frac{f_5}{q^2} + \frac{f_7}{M^2_W}\nonumber\\
b_3 & = & i\frac{f_5}{q^2} + \frac{f_7}{M^2_W}\,.
\end{eqnarray}

To take this comparison one step further, suppose that the effects
of $P$ and $C$ violation are negligible in the $VV\gamma$ coupling.
It then follows from Eqs.\ (\ref{Ptransform}) and (\ref{Ctransform}) 
that the only non-vanishing terms in Eq.\ (\ref{Gammagentranschargdiag}) are
$a'_1$, $a'_2$ and $a_3$.  Apart from a slight change of notation,
this is the form adopted, for example, in Ref.\cite{bardeen,papava}.
%
%
\section{Conclusions}\setcounter{equation}{0}
\label{sec:conclusions}

We have studied in this article the 
structure of the couplings
of spin-1 particles to the photon.  
In Section \ref{sec:parametrization} we considered
the general form that the electromagnetic vertex
can have, consistent with Lorentz invariance,
and we established the physical interpretation
of the various form factors that parametrize the vertex 
in terms of the static electromagnetic moments.
In Section \ref{sec:discretesymmetries} we
derived the consequences of the various
discrete space-time symmetries on the form factors,
paying special attention to the case of neutral bosons
and in particular to the case of self-conjugate bosons.
In the latter case, we derived some results that
are analogous to similar results that are known
to hold regarding the electromagnetic
couplings of Majorana fermions.
Finally, in Section \ref{sec:gaugeinv}
we analyzed in detail the implications due
to gauge invariance for the structure of the vertex
function.  In particular, several results concerning
the electromagnetic properties of self-conjugate bosons
were obtained there.  For example, it was shown that
while a self-conjugate particle cannot 
cannot have any static electromagnetic moments,
in the most general case it can have an
electromagnetic vertex characterized by
two form factors.  This result is analogous
to the corresponding one for Majorana fermions,
which cannot have static electromagnetic moments either,
but it can have an electromagnetic vertex characterized
by an axial charge radius form factor.
For the three-photon vertex, which vanishes in pure QED due
to Furry's theorem, we obtained a general form which
need not vanish due to the breaking of the Charge Conjugation
symmetry by the weak interactions.

We have been motivated by the fact that experimental 
studies of this kind of coupling will be feasible
in the future.  In this context, the
analysis that we have presented can be a useful
in at least two ways.  On one hand,
it can serve as a guide to parametrize any possible
deviation of the couplings from the values predicted
by Standard Model, in a way that is general
and model-independent. On the other hand,  
whenever the study of a new kind of phenomena is accesible
to us, it is useful to keep in mind
that our present knowledge may be shaken
by new discoveries in a  more unexpected way
than simply just a deviation from the
detailed values predicted by the Standard Model
for a given physical quantity.  The results of our
analysis can be used to test deviations
from fundamental physical principles, such as
gauge invariance and crossing symmetry,
in the context of the processes described by the
electromagnetic couplings of vector bosons.

We thank Sonjoy Majumder for discusssions. The research of JFN was 
supported in part by
the US National Science Foundation Grant PHY-9600924.

\end{document}